# Solid phase crystallization under continuous heating: kinetic and microstructure scaling laws

J. Farjas[*] and P. Roura

GRMT, Department of Physics, University of Girona, Campus Montilivi, Edif. PII, E17071 Girona, Catalonia, Spain

*Corresponding author: jordi.farjas@udg.es

**Abstract**

The kinetics and microstructure of solid phase crystallization under continuous heating conditions and random distribution of nuclei are analyzed. An Arrhenius temperature dependence is assumed for both nucleation and growth rates. Under these circumstances, the system has a scaling law such that the behavior of the scaled system is independent of the heating rate. Hence, the kinetics and microstructure obtained at different heating rates only differ in time and length scaling factors. Concerning the kinetics, it is shown that the extended volume evolves with time according to $\alpha_{ex} = [\exp(\kappa C t')]^{m+1}$ where $t'$ is the dimensionless time. This scaled solution not only represents a significant simplification of the system description, it also provides new tools for its analysis. For instance, it has been possible to find an analytical dependence of the final average grain size on the kinetic parameters. Concerning the microstructure, the existence of a length scaling factor has allowed the grain size distribution to be numerically calculated as a function of the kinetic parameters.

## 1. Introduction

Crystallization of amorphous materials and other solid state transformations usually involve random nucleation and growth. Under this assumption, the phase transformation is described by the Kolmogorov-Johnson-Mehl-Avrami theory (KJMA) [1-6]. The transformed fraction, $\alpha$, is related with the extended transformed fraction, $\alpha_{ex}$, through the so-called KJMA relation:

$$\alpha(t) = 1 - \exp[-\alpha_{ex}(t)] \quad . \tag{1}$$

$\alpha_{ex}$ would be the transformed fraction if grains grew through each other and overlapped without mutual interference, i.e.:

$$\alpha_{ex}(t) = \int_0^t I(u) v_{ex}(u,t) du \quad , \tag{2}$$

where $I$ is the nucleation rate per unit volume and $v_{ex}(u,t)$ is the extended volume transformed at time $t$ by a single nucleus created at time $u$

$$v_{ex}(u,t) = \sigma \left( \int_u^t G(z) dz \right)^m \quad . \tag{3}$$

In Eq. (3), $\sigma$ is a shape factor (e.g., $\sigma = 4\pi/3$ for spherical grains), $G$ is the growth rate and $m$ depends on the growth mechanism [7] (e.g., m=3 for three dimensional, 3D, growth).

For the particular case of isothermal transformations, where growth and nucleation rates are constant in time, Eqs. (2)-(3) have an analytical solution:

$$\alpha_{ex}(t) = \left( \frac{\sigma I G^m}{m+1} \right) t^{m+1} \quad . \tag{4}$$

Unfortunately, owing to the dependence of $G$ and $I$ on temperature, general exact solutions do not exist for non-isothermal conditions. Accordingly, a number of published works have developed different theoretical and numerical approaches to analyze non-isothermal phase transformations within the framework of KJMA theory [8-30]. Recently, a quasi-exact solution of the KJMA theory was obtained under continuous heating conditions [31].

A useful approach to investigate the kinetics and grain morphology consists of finding a scaling law such that the system behavior is universal. This method has been successfully used for the isothermal case [32]. In this case the time, $\tau$, and length, $\lambda$, scaling factors are [33]:

$$\tau = \left(I\, G^m\right)^{-1/(m+1)} \quad \text{and} \quad \lambda = \left(\frac{G}{I}\right)^{1/(m+1)} . \tag{5}$$

When time is scaled in Eq. (4), one gets a universal solution (independent of $I$ and $G$):

$$\alpha_{ex}(t') = (\kappa t')^{m+1} , \tag{6}$$

where,

$$\kappa \equiv \sqrt[m+1]{\frac{\sigma}{m+1}} , \tag{7}$$

and $t' \equiv t/\tau$ is the dimensionless time.

In this paper we will show that a similar scaling law applies for transformations at a constant heating rate (Sec. 2). For a given ratio between the activation energies of $I$ and $G$, there exists an approximate scaled solution independent of the heating rate. Accordingly, the kinetics and microstructure for any heating rate can be obtained from this scaled solution simply by multiplying the dimensionless time and length values by the corresponding scaling factors. In Sec. 3 we obtain the scaled solution for the transformation kinetics, $\alpha(t')$, which represents a significant simplification when compared to the quasi-exact solution recently published [31].

Apart from the transformation kinetics it would be very useful to know the resulting material's microstructure because many of the material's physical properties are microstructure-dependent. Surprisingly, work related to the microstructure obtained under continuous heating conditions is very scarce. As far as we know, only Crespo et al. [34] have addressed this problem for a particular case. In Section 4, and thanks to the simplicity of the scaled solution, an analytical expression is obtained for the average grain size. Additionally, we numerically analyze the dependence of the grain size distribution on the ratio between the nucleation and growth activation energies. Finally, in Section 5 the limits of thermally activated nucleation are analyzed. It will be shown that when the activation energies of nucleation and growth are significantly different, the model of pre-existing nuclei is more adequate. A scaled exact solution for pre-existing nuclei is also included.

## 2. The scaling law

In most practical situations where continuous nucleation takes place, it is possible to assume an Arrhenius temperature dependence for both $I$ and $G$ [9-12,21,25,35]:

$$I = I_0 \exp(-E_N/k_B T) \text{ and } G = G_0 \exp(-E_G/k_B T) \quad , \tag{8}$$

where $E_N$ and $E_G$ are the respective activation energies for nucleation and growth, $k_B$ is the Boltzmann constant and $T$ is the temperature. Under this assumption, Eqs. (1)-(3) have a quasi-exact solution [31]:

$$\alpha = 1 - \exp\left\{-\left[k_0 C \frac{E}{\beta k_B} p\left(\frac{E}{k_B T}\right)\right]^{m+1}\right\} \quad , \tag{9}$$

where $k_0 \equiv \kappa\left(I_0 G_0^m\right)^{1/m+1}$, $E \equiv \dfrac{E_N + m E_G}{m+1}$, $C \equiv \left(\dfrac{(m+1)! E^{m+1}}{\prod\limits_{i=0}^{m}(E_N + i E_G)}\right)^{1/m+1}$, $p(x) \equiv \int_x^\infty \dfrac{\exp(-u)}{u^2} du$

and $\beta$ is the constant heating rate; $\beta \equiv dT/dt$. Note that, according to Eqs. (1) and (9), $\alpha_{ex} = \left[k_0 C \dfrac{E}{\beta k_B} p\left(\dfrac{E}{k_B T}\right)\right]^{m+1}$. Moreover, $\alpha(t)$ in Eq. (9) is the exact solution of the non-isothermal KJMA rate equation [31]:

$$\frac{d\alpha}{dt} = (m+1) \cdot C \cdot k(T) \cdot (1-\alpha)[-Ln(1-\alpha)]^{m/m+1} \quad , \tag{10}$$

where $k(T) \equiv k_0 e^{-E/k_B T}$.

The time, $\tau_P$, and length, $\lambda_P$, scaling factors we propose here are inspired by the isothermal case, Eq. (5). Since $I$ and $G$ depend on time through temperature for constant heating, we define the scaling factors using the values of $I$ and $G$ for a particular temperature. A logical choice is the well-defined peak temperature, $T_P$, i.e., the temperature at which the transformation rate is maximum:

$$\tau_P \equiv \left(I\, G^m\right)^{\frac{-1}{m+1}}\bigg|_{T=T_P} = \left(I_0\, G_0^m\right)^{\frac{-1}{m+1}} \exp\left(\frac{E}{k_B T_P}\right)$$

$$\lambda_P \equiv \left(\frac{G}{I}\right)^{\frac{1}{m+1}}\bigg|_{T=T_P} = \left(\frac{G_0}{I_0}\right)^{\frac{1}{m+1}} \exp\left(-\frac{E_G - E_N}{(m+1)k_B T_P}\right)$$

(11)

where $T_P$, is given by Eq. (A2) (see Appendix A).

Under the approximation that the crystallization takes place in a relatively narrow temperature range,

$$\frac{1}{T} - \frac{1}{T_P} \approx \frac{T_P - T}{T_P^2} \quad , \quad (12)$$

the dimensionless growth and nucleation rates become (see Appendix A):

$$G' = \exp\left[\frac{E_G}{E}\kappa C t'\right] \text{ and } I' = \exp\left[\frac{E_N}{E}\kappa C t'\right] \quad , \quad (13)$$

where $t' \equiv t/\tau_P$ is the dimensionless time. Note that the dimensionless growth and nucleation rates do not depend on the heating rate, they only depend on the geometrical factor $\sigma$ and the ratio $E_N/E_G$ through the constants $\kappa, E_G/E$ and $C$, respectively. Hence, for a given ratio $E_N/E_G$ the transformation kinetics and grain morphology for different $\beta$ differ only by the time and length scaling factors $\tau_P$ and $\lambda_P$, respectively. Therefore kinetics and microstructure can be obtained from the scaled system simply by multiplying the dimensionless time and length by $\tau_P$ and $\lambda_P$, respectively.

Equation (9) is obtained under the assumptions that the critical nuclei size, the transformation rate at the initial temperature, $T_0$, and the incubation time for nucleation are negligible. The first assumption relies on the fact that the average grain size is usually much larger than the critical nuclei size. Thus, this approximation only affects the very early stages of crystallization. Concerning the second assumption, it is based on the fact that, in well designed experiments, $T_0$ is low enough to ensure that the experimental results do not depend on $T_0$. Finally, the existence of a finite incubation time would modify Eq. (9). However, as the incubation time is linked to the crystallization kinetics, in many cases an approximate relation equivalent to (A2) is expected and the scaling law is still valid. For instance, we have verified the validity of the scaling law for the case of crystallization of a-Si where the activation energy of the incubation time is similar to that of crystallization [36].

## 3. Scaled approximate solution for the transformation kinetics

In this section, we will find a scaled expression for $\alpha(t)$ (i.e., independent of $\beta$) which virtually coincides with the quasi-exact solution. Let us rewrite the non-isothermal KJMA equation [Eq. (10)] for $\alpha_{ex}$:

$$\frac{d\alpha_{ex}}{dt} = (m+1)k(T)C\alpha_{ex}^{\frac{m}{m+1}} \quad , \tag{14}$$

and show that, under the approximation of Eq.(12), it is scalable with time. With Eq. (12) $k(T)$ becomes:

$$k(T) \approx \frac{\kappa}{\tau_P}\exp\left(\kappa C \frac{t}{\tau_P}\right) \quad . \tag{15}$$

Once $k(T)$ is substituted in Eq. (14), a scaled equation results:

$$\frac{d\alpha_{ex}}{dt'} = (m+1)\kappa C e^{\kappa C t'}\alpha_{ex}^{\frac{m}{m+1}} \quad , \tag{16}$$

where $t' \equiv t/\tau_P$ is the dimensionless time. Integration of Eq. (16) delivers the scaled solution for $\alpha_{ex}$:

$$\alpha_{ex}(t') = \left[\exp(\kappa C t')\right]^{m+1} \quad , \tag{17}$$

after imposing that $\alpha_{ex} = 1$ at the peak temperature [31] (i.e., at $t'=0$). Finally the scaled solution for the transformed fraction is obtained after combining Eqs. (1) and (17):

$$\alpha(t') = 1 - \exp\left(-\left[\exp(\kappa C t')\right]^{m+1}\right) \quad . \tag{18}$$

Alternatively, Eq. (17) can be obtained after integration of Eqs. (2)-(3) once the dimensionless rate constants $G'$ and $I'$ [Eq. (13)] are substituted there.

To verify the validity of the time scaling and the proposed scaled solution for $\alpha(t')$, we have calculated $\alpha(t)$ and the transformation rate, $d\alpha/dt$, from the crystallization of amorphous silicon at two extreme heating rates of $\beta = 0.005$ and 100 K/min which can be considered the lower and upper limits for most experiments. Then, time and transformation rate are scaled for these particular heating rates and compared to the scaled solution $\alpha(t')$ given by Eq. (18). and its derivative $d\alpha'/dt$. The results are plotted in Fig. 1. The calculation of $\alpha(t)$ has been done using a numerical method which delivers the exact solution of Eqs. (1)-(3) [32]. All calculations described in this section have been done for isotropic 3D crystallization ($m = 3$, $\sigma = 4/3\pi$) of amorphous silicon ($I$ and $G$, detailed in Table I).

The coincidence for both heating rates and the scaled solution is excellent. The discrepancies in the transformed fraction for both heating rates are lower than 3 10$^{-4}$ despite the large shift in peak temperatures from 571.6 to 768.7 ºC and the very different time-scaling factors of 1.72 10$^5$ and 13.0 s for $\beta = 0.005$ and 100 K/min, respectively. It is worth noting that, although the two time scales differ by more than five orders of magnitude, the scaling law is still valid. In fact, the usefulness of the approximation made in Eq. (12) is based on the exponential dependence of the growth and nucleation rates on temperature, i.e., the Arrhenius dependence. This strong dependence on temperature limits the crystallization process to a narrow temperature range even when the heating rate is as low as 0.005 K/min.

The accuracy of the scaled solution can be tested for a wide range of $E_N/E_G$ values through the full width at half maximum (FWHM) of the transformation rate peak. From the scaled solution, Eq. (18), the calculation of the FWHM, $\Delta t_{HM}$, is straightforward and results in (see Appendix B):

$$\frac{\Delta t_{HM}}{\tau_P} = \frac{2.44639}{(m+1)\kappa C} \qquad (19)$$

In Fig. 2 we see that this value departs only slightly from the exact one when $E_N$ is very different from $E_G$, the discrepancy being higher for $E_N \ll E_G$.

Let us highlight the formal simplicity of the scaled solution [Eq. (18)] when compared with the quasi-exact solution [Eq. (9)]. This simplicity has been reached without any significant loss of accuracy in the range of transformed fractions of practical interest (say, $0.01 < \alpha < 0.99$).

In contrast with the isothermal case [Eq. (6)] we see that the scaled solution for continuous heating is not universal (independent of $G$ and $I$) but depends on the particular value of the ratio between the activation energies $E_G$ and $E_N$ (through the parameter $C$). This dependence has important consequences for the microstructure development, which will be analyzed in the next section.

## 4. Grain size morphology

In this section we will verify the length scaling law proposed in Sec. 2 [Eq. (11)] and analyze the dependence of the final microstructure on the kinetic parameters.

To characterize the final microstructure we have calculated the grain size distribution, where the grain size of an individual grain, $i$, is defined as:

$$r_i \equiv \sqrt[3]{\frac{3v_i}{4\pi}} \qquad , \qquad (20)$$

and $v_i$ is the actual grain volume. The numerical algorithm used for the calculation of the grain size distribution is described in [32].

To verify the length scaling law we have calculated the grain size distributions from the crystallization of amorphous silicon at two extreme heating rates of $\beta$ =0.005 and 100 K/min. Then the grain size distributions are scaled by dividing the grain size by the length scaling factor $\lambda_P$ [Eq. (11)]. The results are plotted in Fig. 3. The length scaling factors are 1.132 and 0.271 microns for $\beta = 0.005$ and 100 K/min, respectively. The discrepancies between both distributions are within the numerical accuracy of the algorithm. Thus, once scaled, the grain size distributions do not depend on the particular heating rate and merge in one single distribution. Indeed, as explained in Sec. 2 the scaled grain size distribution only depends on the $E_N / E_G$ ratio.

To characterize the grain size distributions from the numerical simulations, we have calculated the average grain size, $<r>$, the mean grain radius, $\bar{r}$ and its standard deviation, $\sigma_r$, defined as:

$$<r> \equiv \sqrt[3]{\frac{1}{N}\sum_{i=1}^{N} r_i^3} \ , \ \bar{r} \equiv \frac{1}{N}\sum_i r_i \ \text{and} \ \sigma_r \equiv \sqrt{\frac{1}{N}\sum_i (r_i - \bar{r})^2} \qquad , \qquad (21)$$

where $N$ is the final number of grains. The average grain size after 3D crystallization as a function of $E_N / E_G$ is reported in Fig 2. The grain size distribution for a series of $E_N / E_G$ ratios below or above unity are plotted in Figs. 4(a) and 4(b), respectively, and, finally, the values of $\bar{r}$ and $\sigma_r$ are detailed in Fig. 5.

*4.1 Analytical solution for the average grain size*

Before looking at the grain size distributions in detail, let us take advantage of the scaled solution for $\alpha$ [Eq. (18)] which allows us to find an analytical expression for the average grain size. According to Ref. [32], $<r>$ can also be calculated from the number of grains formed after complete crystallization:

$$<r> = \sqrt[m]{\frac{1}{\sigma}\frac{V}{N}} \qquad , \qquad (22)$$

where $N$ can be obtained from $\alpha(t)$:

$$N = V\int_0^\infty I(u)(1-\alpha(u))du \quad . \qquad (23)$$

In Appendix B, the scaled solution has been substituted in Eq. (23) and an analytical expression for $<r>$ has been obtained:

$$\frac{<r>}{\lambda_P} = \left( \frac{\kappa^m}{C} \Gamma\left( \frac{E_N}{(m+1)E} \right) \right)^{-\frac{1}{m}} \quad , \quad (24)$$

where $\Gamma$ is the gamma function [37]. It has been plotted in Fig. 2 where it can be compared with the exact values obtained for the two extreme heating rates of 0.005 and 100 K/min. It can be concluded that Eq. (24) gives the average grain size with very good accuracy unless $E_N / E_G \ll 1$, where discrepancy is below 10%.

*4.2 Dependence of the grain size distribution on $E_N / E_G$*

Let us now focus our attention on the grain-size distributions of Fig. 4. When $E_N / E_G = 1$, the distribution coincides with the isothermal distribution obtained in ref. [32]. This is as expected because *G* and *I* have the same temperature dependence and their ratio is constant. Consequently, the temperature has an effect on the rate at which the transformation proceeds but not on the microstructure. In this particular case, the grain size distribution is independent of the thermal history. For $E_N \neq E_G$, the grain-size distribution departs progressively from the isothermal one, and when the ratio $E_N / E_G$ is far from unity, the distributions have characteristic shapes which can be readily understood.

When $E_N / E_G < 1$, during the first stages of the transformation, nucleation dominates over growth. Consequently, the nuclei density is higher when compared to the isothermal case. Thus, when $E_N / E_G$ diminishes, the average grain is reduced (Fig. 3). Concerning the bell-shaped grain-size distribution for $E_N / E_G \ll 1$ [Fig. 4(a)], it can be explained by the fact that most nuclei are formed at a temperature range where they are not allowed to grow significantly. This means that they grow together at higher temperatures leading to a narrow distribution of grain sizes. In Fig. 5 we see that, indeed, the standard deviation diminishes drastically for $E_N / E_G \ll 1$.

In contrast, when $E_N/E_G > 1$ during the first stages of crystallization, growth dominates and the nucleation rate increases progressively as crystallization proceeds. Since the time left for growing is lower for the nuclei that appear later, the density of small grains will be higher than for larger grains [as shown in Fig. 4(b)]. In fact, from Fig. 4(b), one can infer that the slow initial nucleation results in the formation of a small quantity of large grains. Moreover, this initial low nucleation rate results in a reduction of the transformation rate which is manifested in Fig. 2 as a monotonous increase of $\Delta t_{HM}$ with $E_N/E_G$. In addition, nucleation takes place during a longer time interval and, consequently, the grain size distribution contains larger grains as $E_N/E_G$ increases (see Fig. 5).

Finally, it is worth noting that when $E_N/E_G$ increases, the delayed nucleation results in an increased number of phantom nuclei [2]. Some authors have claimed that, in clear disagreement with Avrami's assumption, they must be excluded in the calculation of $\alpha_{ex}$. For the simulations carried out in this work, the ratio between phantom nuclei and 'real' grains increases from 0.11 when $E_N/E_G = 0.1$ to 0.98 when $E_N/E_G = 10$. In all these simulations, the agreement between the transformed fractions calculated from the microstructure and from the numerical solution of Avrami's model [Eq. (1)] is excellent.

**5. Limits of thermally activated nucleation**

From a formal point of view, the analysis given in Sec. 2-4 for continuous nucleation can be applied for any arbitrary value of the ratio $E_N/E_G$. In the following, we will argue that, when this ratio is far from unity, the material will follow the kinetics of pre-existing nuclei (described in Appendix C), when nucleation is not thermally activated but a constant density of nuclei, $n_0$, already exists before they grow. When $E_N \ll E_G$, nucleation takes place early and, eventually, its rate may vanish before the onset of particle growth (site saturated nucleation [21,38]). Consequently the nuclei grow as if they were preexistent to the growth stage. On the other hand, when $E_N \gg E_G$, homogeneous nucleation is less viable. In most practical situations, when $E_N \gg E_G$, nucleation is catalyzed by inclusions and the container walls [35,39,40], i.e., it is virtually impossible to prevent heterogeneous nucleation. In this case, again, one can also apply the model of preexisting nuclei provided that nuclei are randomly distributed [41,42]. In the case of heterogeneous nucleation, the latter condition can be jeopardized by a particular distribution of the external nucleation sites. However, in several practical situations and in the case of heterogeneous nucleation localized at the container walls it is possible to assume that nucleation sites are randomly distributed. Thus, the problem can be solved by assuming an initial surface density of preexisting nuclei [43-45].

A universal scaled solution can also be obtained for the case of pre-existing nuclei (see Appendix C). Calculations, like those done for continuous nucleation in Sec. 2, show that the kinetics is scalable with a similar accuracy ($\Delta\alpha < 4.10^{-4}$ between 0.005 and 100 K/min).

## 6. Conclusions

In this paper, we have shown that, when time and length are properly scaled, the description of solid state crystallization under annealing at a constant heating rate becomes very simple. For a given material, the time dependencies of the transformed fraction obtained at different heating rates merge into one single scaled solution. The accuracy of this simplified kinetics has been tested against exact numerical solutions of the KJMA equations. Within the range of $\alpha$ values of interest ($0.01 < \alpha < 0.99$), the agreement is excellent. Apart from a geometrical parameter, this scaled solution depends only on the particular growth and nucleation rates through one single parameter: the ratio of activation energies, $E_N / E_G$.

In addition to the crystallization kinetics, it has been shown that the grain size distribution can be scaled with a characteristic length. Again, for a given $E_N/E_G$ ratio, the scaled distributions do not depend on the particular heating rate. From the scaled kinetic equation, it has been possible to obtain the analytical dependence of the average grain size on $E_N/E_G$. The grain size obtained after isothermal crystallization coincides with that obtained after continuous heating only when $E_N/E_G = 1$. Although small deviations are predicted for $E_N \ll E_G$, they are probably not high enough to induce important changes in the material's properties.

The scaled distributions have been calculated for a series of $E_N/E_G$ values ranging from 0.1 to 10 with a numerical algorithm which simulates the microstructure development. It has been shown that, for $E_N \ll E_G$, the distribution of grain sizes is quite narrow around the average value whereas, for $E_N \gg E_G$, the density of grains diminishes monotonically as the radius increases.

For the sake of completeness, the kinetics and grain size distribution have been calculated for the case of preexisting nuclei. It has been shown that it is also possible to find appropriate time and length scaling factors.

Finally, our analysis relies on the fact that the transformation is thermally activated and, consequently, that it takes place in a narrow temperature range. Indeed, many real transformations are thermally activated, thus we believe that our approach can by applied to a large number of transformations.

**Acknowledgments**

This work has been supported by the Spanish *Programa Nacional de Materiales* under contract number MAT2006-11144.

**Appendix A. Dimensional scaling law for the case of continuous nucleation**

The time, $\tau_P$, and length, $\lambda_P$, scaling factors are defined in Eq. (11) where the peak temperature, $T_P$, is given by:

$$\left.\frac{d^2\alpha}{dt^2}\right|_{T_P} = 0 \quad . \tag{A1}$$

Substitution of Eqs. (9) and (10) in Eq. (A1) leads to the value of $T_P$ as the solution of an algebraic equation:

$$\frac{\beta}{T_P^2} = \frac{k_0 k_B C}{E} e^{-\frac{E}{k_B T_P}} \tag{A2}$$

The scaled system is universal (independent of $\beta$) provided that the dimensionless growth and nucleation rates do not depend on $\beta$. Actually, the dimensionless growth and nucleation rates are:

$$G' \equiv G\frac{\tau_P}{\lambda_P} = e^{-\frac{E_G}{k_B}\left(\frac{1}{T}-\frac{1}{T_P}\right)} \quad \text{and} \quad I' \equiv I\lambda_P^m \tau_P = e^{-\frac{E_N}{k_B}\left(\frac{1}{T}-\frac{1}{T_P}\right)} \quad . \tag{A3}$$

Unfortunately, the result does depend on $\beta$ through the relationship between $T$ and $t$. To suppress this dependence we will suppose that the temperature range where the crystallization takes place is relatively narrow:

$$\frac{1}{T} - \frac{1}{T_P} = \frac{T_P - T}{T_P^2}\left(1 - \left(\frac{T-T_P}{T}\right) + ...\right) \approx \frac{T_P - T}{T_P^2} \quad . \tag{A4}$$

Furthermore, selecting a time scale requires selecting a scale factor as well as a time origin. This origin must correspond to an equivalent state for any dimensional system (any particular value of $\beta$). Here again the natural choice is the time at which the transformation rate is maximum:

$$T = T_P + \beta t \Leftrightarrow T_P - T = -\beta t \quad . \tag{A5}$$

Then, substitution of Eqs. (A5) and (A2) into Eq. (A4) gives:

$$\frac{1}{T} - \frac{1}{T_P} \approx \frac{-\beta t}{T_P^2} = -\frac{\beta k_0 k_B C}{E} e^{-\frac{E}{k_B T_P}} = -\frac{k_B}{E}\kappa C \frac{t}{\tau_P} \quad . \tag{A6}$$

Thus, the dimensionless growth and nucleation rates become:

$$G' = \exp\left[\frac{E_G}{E}\kappa C t'\right] \text{ and } I' = \exp\left[\frac{E_N}{E}\kappa C t'\right] \quad , \quad \text{(A7)}$$

where $t' \equiv t/\tau_P$ is the dimensionless time.

**Appendix B. Analytical calculation of $<r>$ and $\Delta t_{HM}$ for the scaled system**

The total number of grains $N$ is given by Eq. (23). Combining Eqs. (23), (18) and (13), one gets:

$$N\lambda_P^m = \frac{1}{\kappa C}\int_{-\infty}^{+\infty} e^{\left[\frac{E_N}{E}t'\right]} e^{\left[-(\exp[\kappa C t'])^{m+1}\right]} dt' = \frac{1}{(m+1)\kappa C}\int_0^{+\infty} s^{\frac{E_N}{(m+1)E}-1} e^{-s} ds = \frac{1}{(m+1)\kappa C}\Gamma\left(\frac{E_N}{(m+1)E}\right)$$
(B1)

where $s \equiv e^{mt'}$. Finally, $<r>$ is obtained from substituting Eq. (B1) into Eq. (22):

$$\frac{<r>}{\lambda_P} = \left(\frac{\kappa^m}{C}\Gamma\left(\frac{E_N}{(m+1)E}\right)\right)^{-\frac{1}{m}} \quad . \quad \text{(B2)}$$

For the calculation of $\Delta t_{HM}$ we first calculate the transformation rate from Eq. (18):

$$\frac{d\alpha(t')}{dt'} = (m+1)\kappa C \exp\left[-(\exp[\kappa C t'])^{m+1}\right](\exp[\kappa C t'])^{m+1} \quad , \quad \text{(B3)}$$

and the transformation at the maximum is:

$$\left.\frac{d\alpha(t')}{dt'}\right|_{t'=0} = (m+1)\,\kappa C e^{-1} \quad , \quad \text{(B4)}$$

consequently,

$$\frac{\Delta t_{HM}}{\tau_P} = t_2' - t_1' \quad , \quad \text{(B5)}$$

where

$$\left.\frac{d\alpha(t')}{dt'}\right|_{t_1'} = \left.\frac{d\alpha(t')}{dt'}\right|_{t_2'} = \frac{1}{2}\left.\frac{d\alpha(t')}{dt'}\right|_{t'=0} = \frac{1}{2}(m+1)\,\kappa C e^{-1} \quad , \quad \text{(B6)}$$

substituting Eq. (B6) into (B3) one gets

$$-\frac{e^{-1}}{2} = e^{-x}(-x), \quad x \equiv (\exp[\kappa C t'])^{m+1} \quad . \quad \text{(B7)}$$

Equation (B7) has two solutions: $x_1 = 2.67835$ and $x_2 = 0.231961$. By substituting these solutions and Eq. (B2) one obtains:

$$\frac{\Delta t_{HM}}{\tau_P} = \frac{1}{(m+1)\kappa C} \ln \frac{x_2}{x_1} = \frac{2.44639}{(m+1)\kappa C} \quad . \quad (B8)$$

**Appendix C. Universal scaled solution for the case of pre-existing nuclei**

When nucleation is completed prior to crystal growth, the kinetics of the transformation is simpler because it is exclusively governed by the growth rate:

$$\alpha_{ex} = \left[ k'_0 \frac{E_G}{\beta k_B} p\left(\frac{E_G}{k_B T}\right) \right]^m \quad , \quad (C1)$$

where $k'_0 \equiv (\sigma n_0 G_0^m)^{1/m}$ and $n_0$ is the pre-existing nuclei density. Then, the corresponding peak temperature is given by:

$$\frac{\beta}{T_P^2} = \frac{k'_0 k_B}{E_G} e^{-\frac{E_G}{k_B T_P}} \quad . \quad (C2)$$

On the other hand, according to the scaling law for the isothermal case [32,46], the time and length scaling factors are defined as:

$$\tau'_P = \left(n_0 G^m\right)^{-1/m}\bigg|_{T=T_P} = \left(n_0 G_0^m\right)^{-1/m} e^{\frac{E_G}{k_B T_P}} \quad \text{and} \quad \lambda'_P = \left(\frac{1}{n_0}\right)^{1/m} \quad , \quad (C3)$$

and the dimensionless growth rate and nucleation density are:

$$G' \equiv G \frac{\tau'_P}{\lambda'_P} = e^{-\frac{E_G}{k_B}\left(\frac{1}{T} - \frac{1}{T_P}\right)} \quad \text{and} \quad n'_0 \equiv I \lambda'^3_P = 1 \quad . \quad (C4)$$

Supposing again that the temperature range where the crystallization takes places is relatively narrow, one gets

$$G' = \exp\left[\sqrt[m]{\sigma} \cdot t'\right] \quad , \quad (C5)$$

and a much simpler expression results for $\alpha_{ex}$:

$$\alpha_{ex} = \left(\exp\left[\sqrt[m]{\sigma} \cdot t'\right]\right)^m \quad . \quad (C6)$$

For pre-existing nuclei, the grain-size distribution $f(r)$ coincides with the distribution obtained under isothermal conditions. In [32] it has been shown that, for 3D, it can be fitted to a Gaussian distribution (the square correlation coefficient is 0.9998):

$$f(r) = \frac{1}{\sqrt{2\pi}\sigma} e^{-\frac{(r-\mu)^2}{2\sigma^2}} \quad , \tag{C7}$$

where $\mu = 0.6093\lambda_P'$ and $\sigma = 0.0892\lambda_P'$.

For the case of pre-existing nuclei $<r>$ is obtained directly from Eq. (22):

$$\frac{<r>}{\lambda_P'} = \frac{1}{\sqrt[m]{\sigma}} \quad . \tag{C8}$$

For the calculation of $\Delta t_{HM}$ we follow the same procedure developed in Appendix B for continuous nucleation. First we calculate the transformation rate from Eqs. (1) and (C6):

$$\frac{d\alpha(t')}{dt'} = (m-1)\sqrt[m]{\sigma}\exp\left[-\left(\exp\left[\sqrt[m]{\sigma}\cdot t'\right]\right)^m\right]\left(\exp\left[\sqrt[m]{\sigma}\cdot t'\right]\right)^m \quad . \tag{C9}$$

The transformation rate at the maximum is ($t'=0$):

$$\left.\frac{d\alpha(t')}{dt'}\right|_{t'=0} = m\sqrt[m]{\sigma}e^{-1} \quad , \tag{C10}$$

and

$$\left.\frac{d\alpha(t')}{dt'}\right|_{t_1'} = \left.\frac{d\alpha(t')}{dt'}\right|_{t_2'} = \frac{1}{2}\left.\frac{d\alpha(t')}{dt'}\right|_{t'=0} = \frac{1}{2}m\sqrt[m]{\sigma}e^{-1} \quad , \tag{C11}$$

substituting Eq. (C11) into (C9) one gets

$$-\frac{e^{-1}}{2} = e^{-x}(-x), \quad x \equiv \left(\exp\left[\sqrt[m]{\sigma}\cdot t'\right]\right)^m \quad . \tag{C12}$$

Equations (C12) and (B7) are identical so they have the same solutions. By substituting these solutions one obtains:

$$\frac{\Delta t_{HM}}{\tau_P'} = \frac{1}{m\sqrt[m]{\sigma}}\ln\frac{x_2}{x_1} = \frac{2.44639}{m}\frac{1}{\sqrt[m]{\sigma}} \quad . \tag{C13}$$

**Table I**. Experimental parameters of amorphous silicon nucleation and growth rates [36].

| | | |
|---|---|---|
| Nucleation | Activation energy, $E_N$ | 5.3 eV |
| | Preexponential term, $I_0$ | $1.7 \cdot 10^{44}$ s$^{-1}$ m$^{-3}$ |
| Growth | Activation energy, $E_G$ | 3.1 eV |
| | Preexponential term, $G_0$ | $2.1 \cdot 10^7$ s$^{-1}$ m |

# References

[1] A.N. Kolmogorov: On the Statistical Theory of Metal Crystallisation. Izv. Akad. Nauk. SSSR, Ser. Fiz. 1, 355 (1937).

[2] M.J. Avrami: Kinetics of phase change. I General theory. Chem. Phys. 7, 1103 (1939).

[3] M.J. Avrami: Kinetics of Phase Change. II Transformation-Time Relations for Random Distribution of Nuclei. Chem. Phys. 8, 212 (1940).

[4] M.J. Avrami: Granulation, Phase Change, and Microstructure - Kinetics of Phase Change. III. Chem. Phys. 9, 177 (1941).

[5] W.A. Johnson and R.F. Mehl: Reaction kinetics in processes of nucleation and growth. Trans. Amer. Inst. Min. Met. Eng. 135, 416 (1939).

[6] A.A. Burbelko, E. Fraś and W. Kapturkiewicz: About Kolmogorov's statistical theory of phase transformation. Mater. Sci. Eng. A 413-414, 429 (2005).

[7] D.W. Henderson: Experimental-analysis of nonisothermal transformations involving nucleation and growth. J. Thermal Anal. 15, 325 (1979).

[8] T. Ozawa: Kinetics of non-isothermal crystallization. Polymer 12,150 (1971).

[9] D.W. Henderson: Thermal analysis of nonisothermal transformations involving nucleation and growth. J. Non-Cryst. Solids 30, 301 (1979).

[10] T.J.W. Bruijn, W.A. Jong and P.J. Berg: Kinetic parameters in Avrami—Erofeev type reactions from isothermal and non-isothermal experiments. Thermochim. Acta 45, 315 (1981).

[11] H. Yinnon and D.R. Uhlmann: Applications of thermoanalytical techniques to the study of crystallization kinetics in glass-forming liquids .1 Theory. J. Non-Cryst. Solids 54, 253 (1983).

[12] E.J. Mittemeijer: Analysis of the kinetics of phase-transformations. J. Mater Sci. 27, 3977 (1992).

[13] M.C. Weinberg: Nonisothermal surface nucleated transformation kinetics. J. Non-Cryst. Solids 151, 81 (1992).

[14] E. Woldt: The relationship between isothermal and nonisothermal description of Johnson-Mehl-Avrami-Kolmogorov kinetics. J. Phys. Chem. Solids 53, 521 (1992).

[15] M.C. Weinberg: Glass-formation and crystallization kinetics. Thermochim. Acta 280, 63 (1996).

[16] P. Krüger: On the relation between nonisothermal and isothermal Kolmogorov-Johnson-Mehl-Avrami crystallization kinetics. J. Phys. Chem. Solids 54, 1549 (1993).

[17] Y. Long, R.A. Shanks and R.A. Stachurski: Kinetics of polymer crystallization. Prog. Polym. Sci. 20, 651 (1995).

[18] J. Vázquez, C. Wagner, P. Villares and R. Jimenez-Garay R: A theoretical method for determining the crystallized fraction and kinetic parameters by DSC, using non-isothermal techniques. Acta Mater. 44, 4807 (1996).

[19] M.J. Starink and A.M. Zahra: An analysis method for nucleation and growth controlled reactions at constant heating rate. Thermochim. Acta 292, 159 (1997).

[20] V.I. Tkatch, A.I. Limanovskii and V. Yu Kameneva: Studies of crystallization kinetics of $Fe_{40}Ni_{40}P_{14}B_6$ and $Fe_{80}B_{20}$ metallic glasses under non-isothermal conditions. J. Mater Sci. 32, 5669 (1997).

[21] J. Ribeiro-Frade: Crystallization with variable temperature: Corrections for the activation energy. J. Am. Ceram. Soc. 81, 2654 (1998).

[22] P.L. López-Alemany, J. Vázquez, P. Villares and R. Jimenez-Garay: Theoretical analysis on the mechanism and transformation kinetics under non-isothermal conditions - Application to the crystallization of the semiconducting $Sb_{0.06}As_{0.36}Se_{0.48}$ alloy. Mater. Chem. Phys. 65, 150 (2000).

[23] J. Málek: Kinetic analysis of crystallization processes in amorphous materials. Thermochim. Acta 355, 239 (2000).

[24] G. Ruitenberg, E. Woldt and A.K. Petford-Long AK: Comparing the Johnson-Mehl-Avrami-Kolmogorov equations for isothermal and linear heating conditions. Thermochim. Acta 378, 97 (2001).

[25] A.T.W Kempen, F. Sommer and E.J. Mittemeijer: Determination and interpretation of isothermal and non-isothermal transformation kinetics; the effective activation energies in terms of nucleation and growth. J. Mat. Sci. 37, 1321 (2002).

[26] S. Jun, H. Zhang, J. Bechhoefer: Nucleation and growth in one dimension. I. The generalized Kolmogorov-Johnson-Mehl-Avrami model. J. Phys. Rev. E 71,011908 (2005).

[27] F. Liu, F. Sommer and E.J. Mittemeijer: An analytical model for isothermal and isochronal transformation kinetics. J. Mater Sci. 39, 1621 (2004).

[28] F. Liu, F. Sommer and E.J. Mittemeijer: Determination of nucleation and growth mechanisms of the crystallization of amorphous alloys; application to calorimetric data. Acta Mater. 52, 3207 (2004).

[29] P.R. Rios: Relationship between non-isothermal transformation curves and isothermal and non-isothermal kinetics. Acta Mater. 53, 4893 (2005).

[30] J. Vázquez, D. García-Barreda, P.L. López-Alemany, P. Villares and R. Jimenez-Garay R: A study on non-isothermal transformation kinetics - Application to the crystallization of the $Ge_{0.18}Sb_{0.23}Se_{0.59}$ glassy alloy. Mater. Chem. Phys. 96, 107 (2006).

[31] J. Farjas and P. Roura: Modification of the Kolmogorov-Johnson-Mehl-Avrami rate equation for non-isothermal experiments and its analytical solution. Acta Mater. 54, 5573 (2006).

[32] J. Farjas and P. Roura: Numerical model of solid phase transformations governed by nucleation and growth: Microstructure development during isothermal crystallization. Phys. Rev. B 75, 184112 (2007).

[33] J.D. Axe, Y. Yamada: Scaling relations for grain autocorrelation functions during nucleation and growth. Phys. Rev. B 34, 1599 (1986).

[34] D. Crespo and T. Pradell: Evaluation of time-dependent grain-size populations for nucleation and growth kinetics. Phys. Rev. B 54, 3101 (1996).

[35] J.W. Christian. The theory of transformation in metals and alloys, part I, 3rd ed. (Elsevier Science Ltd., Oxford, England, 2002), pp. 423-527.

[36] C. Spinella, S. Lombardo and F. Priolo: Crystal grain nucleation in amorphous silicon. J. Appl. Phys. 84, 5383 (1998).

[37] P.J. Davis. Gamma function and related funcitons, in Handbook of mathematical functions, edited by M. Abramowitz and I.A. Stegun (Dover Publications Inc., New York, NY, 1972) pp. 253.

[38] J.W. Cahn: Transformation kinetics during continuous cooling. Acta Metall. 4, 572 (1956).

[39] D. Turnbull: Formation of crystal nuclei in liquid metals. J. Appl. Phys. 21, 1022 (1950).

[40] J.W. Cahn: The kinetics of grain boundary nucleated reactions. Acta Metall. 4, 449 (1956).

[41] M.P. Shepilov and D.S. Baik: Computer-simulation of crystallization kinetics for the model with simultaneous nucleation of randomly-oriented ellipsoidal crystals. J. Non-Cryst. Solids 171,141 (1994).

[42] V.Z. Belenkii: Generalization of the Kolmogorov model for the crystallization in a bounded space with an arbitrary region of nucleation. Dokl. Akad. Nauk. SSSR 278, 874 (1984).

[43] M.C. Weinberg and R. Kapral: Phase-transformation kinetics in finite inhomogeneously nucleated systems. J. Chem. Phys. 91, 7146 (1989).

[44] M.C. Weinberg: Transformation kinetics of particles with surface and bulk nucleation. J. Non-Cryst. Solids 142, 126 (1992).

[45] J. Vázquez, D. García-Barreda, P.L. López-Alemany, P. Villares and R. Jimenez-Garay: A comparative study on the single-scan and multiple-scan techniques in differential scanning calorimetry Application to the crystallization of the semiconducting $Ge_{0.13}Sb_{0.23}Se_{0.64}$ alloy. Thermochim. Acta 430, 173 (2005).

[46] E. Pineda and D. Crespo: Microstructure development in Kolmogorov, Johnson-Mehl, and Avrami nucleation and growth kinetics. Phys. Rev. B 60, 3104 (1999).

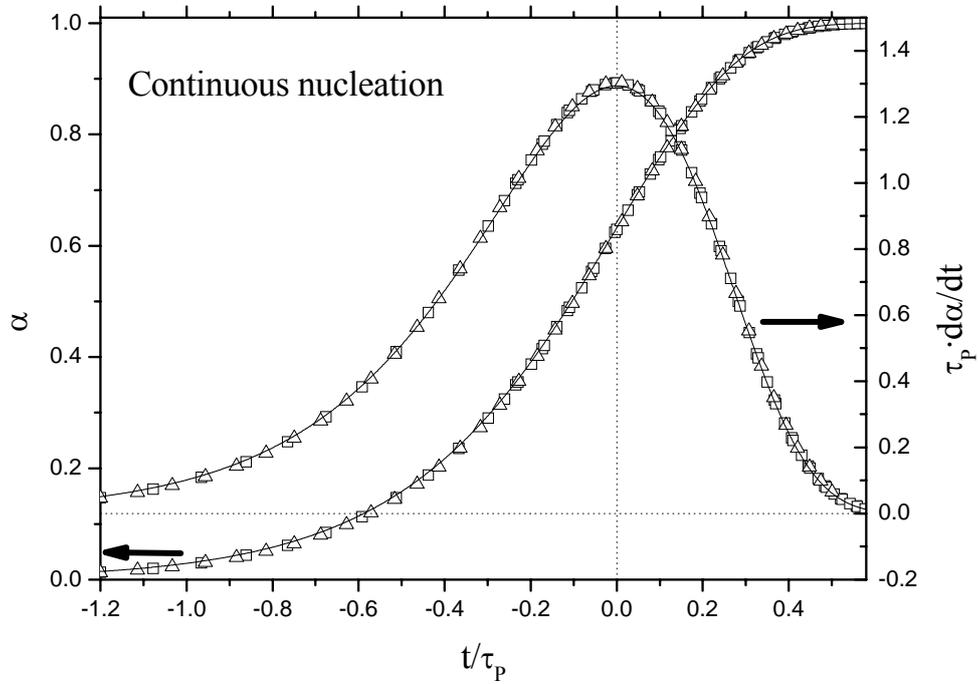

Figure 1. Transformed fraction and transformation rate versus time for 3D crystallization of amorphous silicon under continuous heating. Heating rates: 0.005 K/min (squares) and 100 K/min (triangles). Time and transformation rates have been scaled according to Eq. (11). The solid line is the solution of the scaled system Eq. (18).

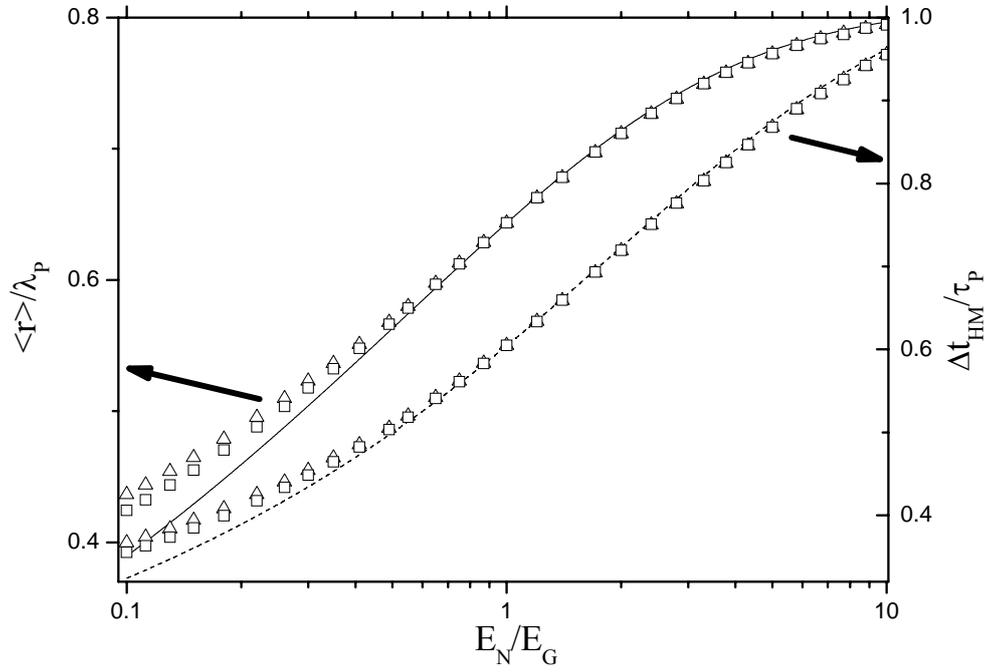

Figure 2. Average grain size, $<r>$, and the FWHM of the transformation rate evolution, $\Delta t_{HM}$. The solid and dashed lines have been calculated from the analytical solution of the scaled system, Eqs. (24) and (19) respectively. The discrete points are the result of a simulation of the crystallization process and can be considered exact (squares: 0.005 K/min; triangles: 100 K/min).

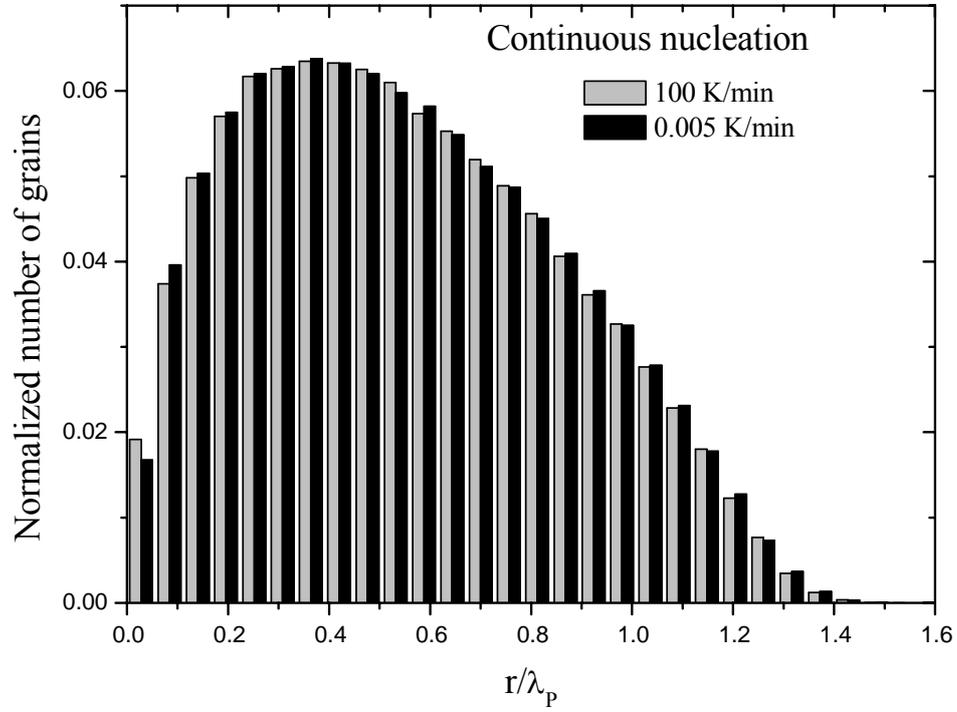

Figure 3. Final grain radius distribution for 3D crystallization of amorphous silicon under continuous heating. Heating rates: 0.005 K/min (black bars) and 100 K/min (grey bars). The radius has been scaled according to Eq. (11).

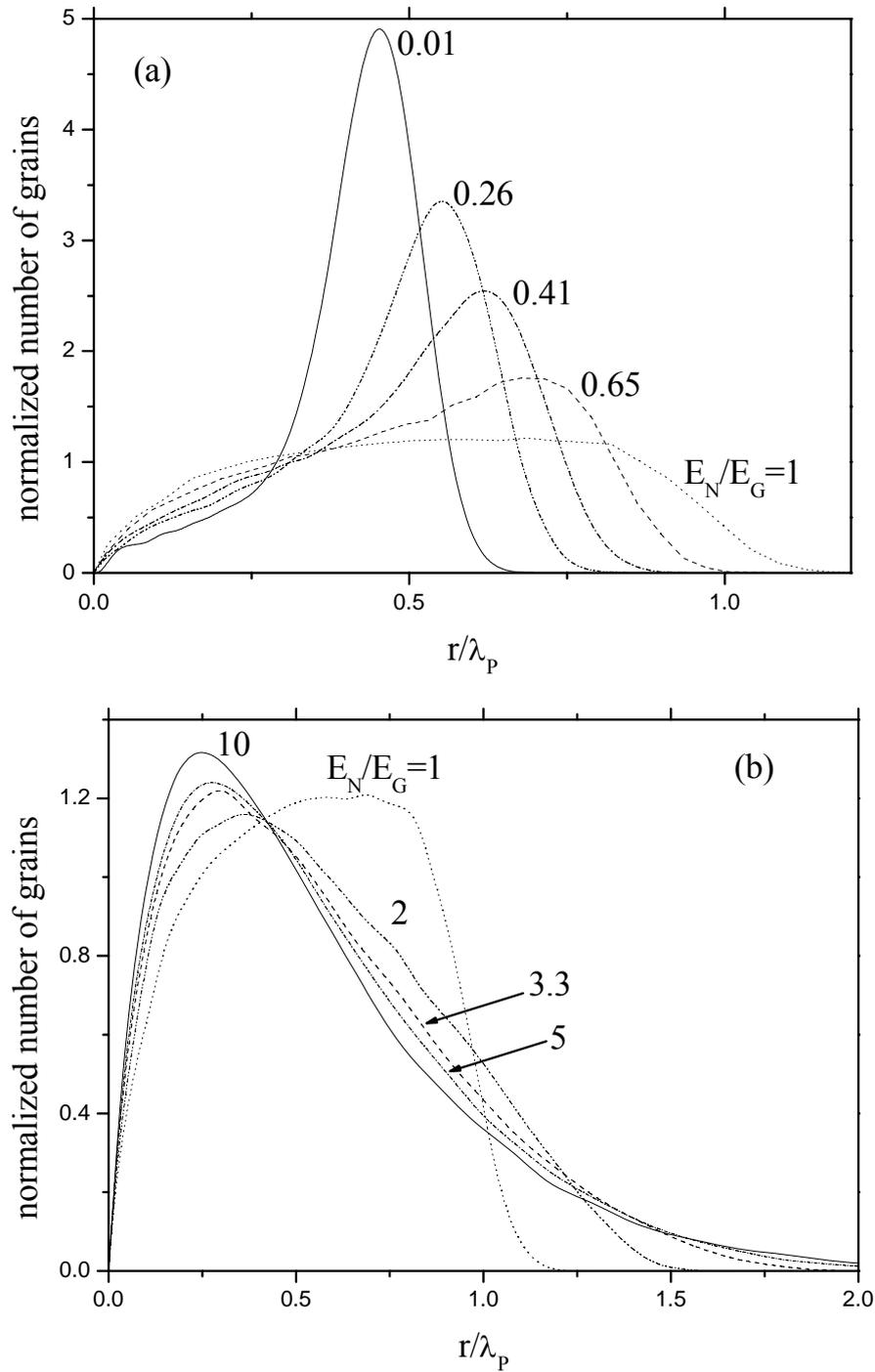

Figure 4. Final grain radius distributions for several values of $E_N / E_G$ resulting from the simulation of the crystallization process at 100 K/min. Owing to the length scaling law, these scaled distributions are almost independent of the heating rate.

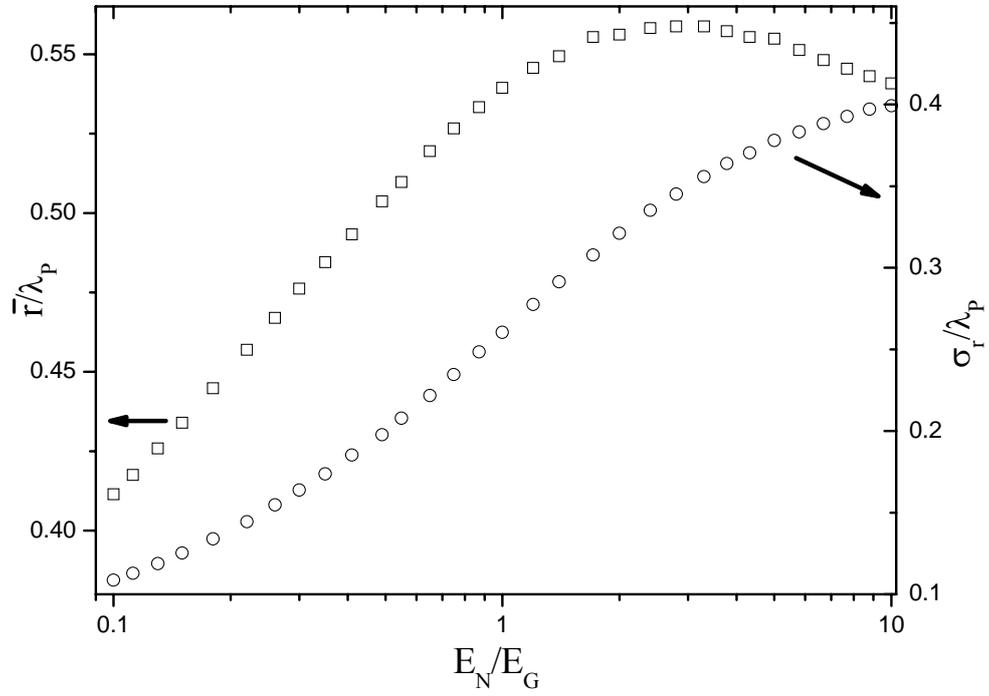

Figure 5. Mean grain radius, $\bar{r}$, and its standard deviation, $\sigma_r$, versus $E_N/E_G$ calculated from the grain size distributions of Fig. 4.